\documentclass[preprint,12pt]{elsarticle}

\usepackage{amssymb}
\usepackage{amsmath}
\usepackage{graphicx}
\usepackage{subcaption}

\usepackage{lineno}

\begin{document}

\begin{frontmatter}



\title{Scale-Invariant Structures of Spiral Waves}

\author{Daniel Sohn\fnref{label1}}
\author{Konstantinos Aronis\fnref{label1}}
\author{Hiroshi Ashikaga\fnref{label1,label2}\corref{cor1}}
\address[label1]{Cardiac Arrhythmia Service, Johns Hopkins University School of Medicine, 600 N Wolfe Street, Carnegie 568, Baltimore, Maryland, USA}
\address[label2]{IHU Liryc, Electrophysiology and Heart Modeling Institute, Fondation Bordeaux Universit\'{e}, F-33600 Pessac-Bordeaux, France}
\cortext[cor1]{To whom correspondence should be addressed.}
\ead{hashika1@jhmi.edu}
\ead[url]{https://www.hiroshiashikaga.org/}

\begin{abstract}
\textit{Background}: Although rotors are thought to be one of the potential mechanisms that maintains spiral waves, the outcome of rotor ablation therapy has been disappointing. To improve our understanding of the mechanism of spiral waves, we developed a multi-scale approach to quantifying the complex interactions among cardiac components as the organizing manifolds of information flow.\\
\textit{Method}: We described rotors in a numerical model of two-dimensional cardiac excitation in a series of spatiotemporal scales by generating a renormalization group, and quantified the Lagrangian coherent structures (LCS) of information flow. To quantify the scale-invariant structures, we compared the value of finite-time Lyapunov exponent (FTLE) between the two corresponding components of the lattice in each spatiotemporal scale of the renormalization group with that of the original microscopic scale.\\
\textit{Results}: Both the repelling and attracting LCS changed across the different spatial and temporal scales of the renormalization group. However, despite the change across the scales, some LCS were scale-invariant. The patterns of those scale-invariant structures were not obvious from the trajectory of the rotors based on the traditional voltage mapping of the cardiac system.\\
\textit{Conclusions}: The Lagrangian coherent structures of information flow underlying spiral waves are preserved across multiple spatiotemporal scales. 
\end{abstract}


\end{frontmatter}


\section{Introduction}
Rotors of the spiral waves are one of the potential mechanisms that maintains fibrillation in human \cite{Lip2016aa}. However, clinical studies to target rotors with catheter ablation therapy demonstrate no significant benefit over the standard ablation approach \cite{haissaguerre2014driver,benharash2015quantitative,gianni2016acute,berntsen2016focal,buch2016long}. Those negative clinical results suggest that our current understanding of the mechanism of fibrillation is far from complete.

Rotors are a macroscopic collective behavior of the heart resulting from interactions among a large number of cells at the microscopic scale. Recently, we have developed an information-theoretic approach to numerically quantifying the complex interactions among cardiac components as the organizing manifolds of information flow \cite{ashikaga2017hidden}. Those manifolds, called the Lagrangian coherent structures (LCS) \cite{haller2000lagrangian,shadden2005definition,haller2015lagrangian}, include two types. Repelling LCS quantify stretching along a material line, and indicate walls through which the flow does not traverse. Attracting LCS quantify folding along a material line, and indicate channels through which flow is funneled. Importantly, some repelling and attracting LCS of rotors become more clearly defined over a longer observation period. This finding suggests that the LCS that is relatively stable over time contributes to maintenance of spiral waves. 

The aim of the study was to describe the complex interactions among cardiac components during spiral waves at multiple scales. To accomplish the aim, we applied iterated coarse-graining and rescaling \cite{kadanoff1966scaling} to the microscopic description of the cardiac system with spiral waves to generate a renormalization group in a series of spatiotemporal scales \cite{ashikaga2018causal,ashikaga2018inter}. We hypothesized that the LCS of information flow underlying spiral waves are scale-invariant.

\section{Methods}
We performed the simulation and the data analysis using Matlab R2017a (Mathworks, Inc.).

\subsection{Model of spiral waves.}
\begin{figure}
  \centering
  \includegraphics[width=0.8\linewidth,trim={0cm 0cm 0cm 0cm},clip]{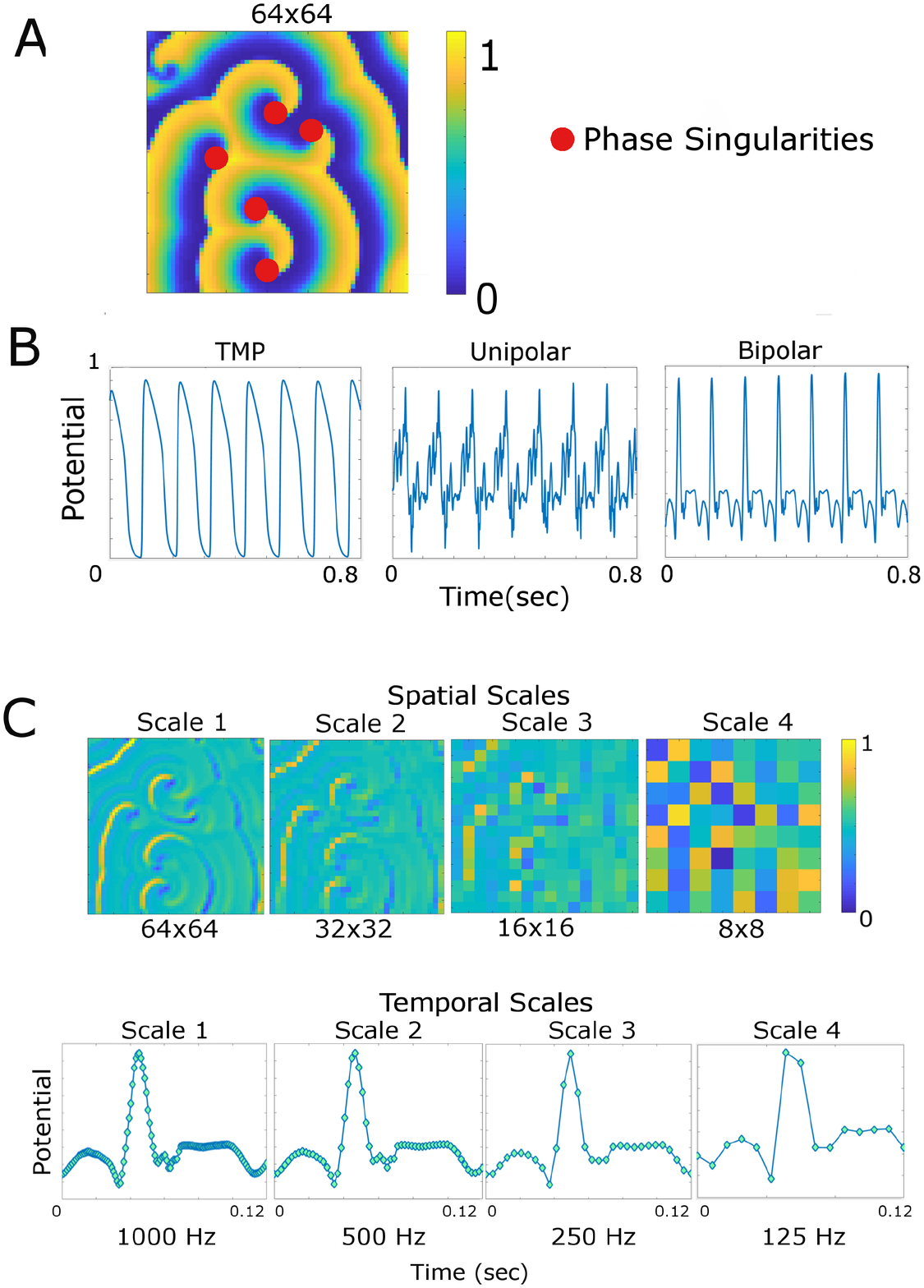}
  \caption{
    \textbf{Conceptual overview.} \textit{A. Spiral waves and rotors.} Spiral waves in a $64 \times 64$ lattice. The transmembrane potential is color-coded (arbitrary unit between 0 and 1). The rotors are defined as the phase singularities of the phase map (filled red circles). \textit{B. Conversion to bipolar signals.} The transmembrane potential (TMP, left panel) derived from the model was converted to unipolar (middle panel) and subsequently to bipolar signals (right panel). \textit{C. Renormalization group.} The bipolar signal is color-coded (arbitrary unit between 0 and 1). Spatial scales include scale 1 ($64 \times 64$ lattice), scale 2 ($32 \times 32$ lattice), scale 3 ($16 \times 16$ lattice), and scale 4 ($8 \times 8$ lattice). Temporal scales include scale 1 (1,000 Hz), scale 2 (500 Hz), scale 3 (250 Hz), and scale 4 (125 Hz). Each circle represents a data sampling point.
  }
  \label{fig:concept}
\end{figure}

We used a deterministic, phenomenological model of the cardiac action potential described by Fenton and Karma \cite{fenton1998vortex}. We chose this model because it accurately reproduces the critical properties of the cardiac action potential to test our hypothesis, such as restitution properties, APD alternans, conduction block, and spiral wave initiation \cite{fenton2002multiple}. The model consists of three variables: the transmembrane potential $V$, a fast ionic gate $u$, and a slow ionic gate $w$.
\begin{equation}
  \frac{\partial V}{\partial t} = \nabla\cdot (D\nabla V) - \frac{I_{fi}+I_{so}+I_{si}+I_{ex}}{C_m}
  \label{eq:FK01}
\end{equation}
Here $C_m$ is the membrane capacitance (= 1 $\mu F/cm^2$), and $D$ is the diffusion tensor, which is a diagonal matrix whose diagonal and off-diagonal elements are 0.001 cm$^2$/msec and 0 cm$^2$/msec, respectively, to represent a two-dimensional (2-D) isotropic system \cite{fenton2002multiple}. The current $I_{fi}$ is a fast inward inactivation current used to depolarize the membrane when an excitation above threshold is induced. The current $I_{so}$ is a slow, time-independent rectifying outward current used to repolarize the membrane back to the resting potential. The current $I_{si}$ is a slow inward inactivation current used to balance $I_{so}$ and to produce the observed plateau in the action potential. $I_{ex}$ is the external current \cite{pertsov1993spiral}. We generated a set of a 2-D $512 \times 512$ isotropic lattice of components by inducing spiral waves with a cross-field stimulation. We spatially coarse-grained the original $512 \times 512$ lattice of time series to a $64 \times 64$ lattice by extracting the top left corner of each  $8 \times 8$ block (Figure \ref{fig:concept}A). In each component, we computed the time series of transmembrane potential for 10 seconds excluding the stimulation period with a time step of 0.1 msec, which was subsequently downsampled at a sampling frequency of 1,000 Hz \cite{ashikaga2018locating}. The rotors were defined as the phase singularities of the phase map as described previously \cite{ashikaga2018causal}. To make our work clinically applicable, the transmembrane potential derived from the model was converted to unipolar and subsequently to bipolar signals (Figure \ref{fig:concept}B) \cite{aronis2018impact}.

\subsection{Renormalization group.}
We generated a renormalization group of the system by a series of spatial and temporal transformation including coarse-graining and rescaling of the original microscopic description of the system (Figure \ref{fig:concept}C) \cite{ashikaga2018causal}. We coarse-grained the system spatially and temporally with decimation by a factor of 2. Spatial decimation transforms a $n \times n$ lattice into a $\frac{n}{2} \times \frac{n}{2}$ lattice by extracting the top left component of each $2 \times 2$ block. Temporal decimation downsampled the time series of each component by a factor of 2. Using a combination of iterative coarse-graining in spatial and temporal axes we created a renormalization group of a total of 16 spatiotemporal scales of the system. The renormalization group included spatial scale 1 ($64 \times 64$ lattice), 2 ($32 \times 32$ lattice), 3 ($16 \times 16$ lattice), and 4 ($8 \times 8$ lattice), and temporal scales 1 (1,000 Hz), 2 (500 Hz), 3 (250 Hz), and 4 (125 Hz).

\subsection{Lagrangian coherent structures of information flow.}
We quantified the Lagrangian coherent structures (LCS) of information flow underlying spiral waves as described previously \cite{ashikaga2017hidden}. Briefly, transfer entropy \cite{schreiber2000measuring} is a non-parametric statistic measuring the directed reduction in uncertainty in one time-series process $X$ (source) given another process $Y$ (destination). 
\begin{align}
T_{X \rightarrow Y}
  &= \sum p(y_{t+1},y^l_t,x^k_t) \log_2 \frac{p(y_{t+1}|y^l_t,x^k_t)}{p(y_{t+1}|y_t^l)} \\
  &= H(y_{t+1}|y^l_t) - H(y_{t+1}|y^l_t,x^k_t)
  ~,
  \label{eq:Txy}
\end{align}
where $k$ and $l$ denote the length of time series in the processes $X$ and $Y$, respectively:
\begin{align}
  x^k_t &= (x_t,x_{t-1},...,x_{t-k+1}) \\
  y^l_t &= (y_t,y_{t-1},...,y_{t-l+1})
  ~.
  \label{eq:xkyl}
\end{align}
In this study we defined $k$ and $l$ such that $x^k_t$ and $y^l_t$ contain a unit time (= 1 sec) of the time-series preceding time $t$ ($k=l$). We used the continuous transfer entropy calculator of the Java Information Dynamics Toolkit (JIDT) to calculate transfer entropy \cite{lizier2014jidt}. We expanded transfer entropy to a vector field on a 2-D lattice. This allowed us to define information flow and velocity as vector quantities at each component, and to visualize information flow and velocity as time-dependent vector fields. Then we defined an information particle as a point that moves with the local information velocity. This allowed us to compute information transport in a Lagrangian perspective. We calculated the finite-time Lyapunov exponent (FTLE) to quantify the spatially- and temporally-localized divergence of trajectories. The forward FTLE field was obtained by integrating the velocity forward in time from $t_0$ to $t$. Similarly, the backward FTLE field was obtained by integrating the negative velocity field backward in time from $t$ to $t_0$. The LCS is defined as ridges, or lines of local maxima, of the FTLE field \cite{shadden2005definition}. A sharp ridge is characterized by a high negative curvature, i.e. high negative eigenvalues $\mu$ of the Hessian matrix of the FTLE field $\nabla^2\mathbf{\Lambda}^t_{t_0}(\mathbf{x}_0)$. For points on the ridge, the gradient of the FTLE field $\nabla\mathbf{\Lambda}^t_{t_0}(\mathbf{x}_0)$ is tangent to the ridge line and perpendicular to the eigenvector $\eta$ corresponding to the smallest eigenvalues $\mu_{\mathrm{min}} < 0$ of the Hessian, which leads to the condition:
\begin{align}
  \eta \cdot \nabla \mathbf{\Lambda}^t_{t_0}(\mathbf{x}_0) = 0
  ~.
  \label{eq:lcs}
\end{align}
Repelling and attracting LCS are derived from the forward and the backward FTLE fields, respectively. The FTLE in te renormalization group was interpolated back to a $64 \times 64$ lattice using cubic spline to allow inter-scale comparison between components.

\subsection{Assessment of scale-invariance of Lagrangian coherent structures.}
The FTLE values of each component of the lattice in each spatiotemporal scale of the renormalization group was compared with that of the original $64 \times 64$ lattice. The difference squared in value between the two corresponding components was calculated and extracted to create another 64x64 lattice that represents the FTLE error. For each component, the error was discretized to 1 when it was less than an arbitrary threshold of 0.1 or 0 when it was equal to or greater than the threshold.

\section{Results}
\subsection{Lagrangian coherent structures of the renormalization group}
\begin{figure}
  \centering
  \includegraphics[width=0.8\linewidth,trim={0cm 0cm 0cm 0cm},clip]{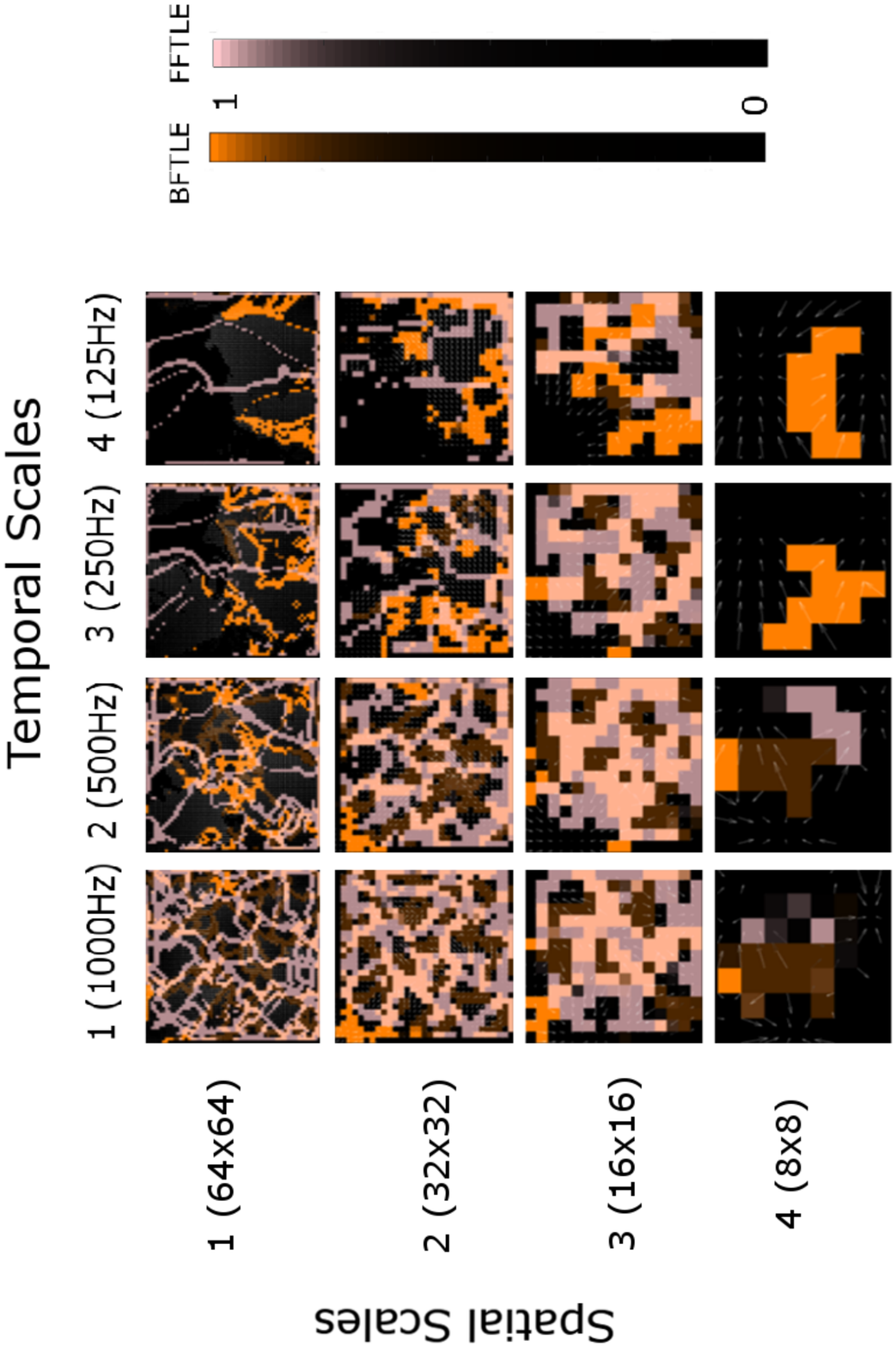}
  \caption{
    \textbf{Lagrangian coherent structures.} The renormalization group of a total of 16 (= $4 \times 4$) spatiotempral scales. The $x$-axis is the temporal scale (1 through 4) and the $y$-axis is the spatial scale (1 through 4). The components shown in orange indicate the repelling Lagrangian coherent structures (LCS) derived from the backward finite-time Lyapunov exponents (BFTLE). The components shown in pink indicate the attracting LCS derived from the forward finite-time Lyapunov exponents (FFTLE). White arrows indicate the information vector field.
  }
  \label{fig:lcs}
\end{figure}

The attracting and repelling LCS in the renormalization group are shown in Figure \ref{fig:lcs}. The original scale (top left panel; spatial scale = 1; temporal scale = 1) shows that both the repelling and attracting LCS cover most of the lattice like a spider web. In the temporal scale 1 (left column), those web-like LCS were preserved in the spatial scales 2 and 3. However, both the repelling and attracting LCS became localized in the spatial scale 4 (left bottom panel). In the temporal scale 2 (second left column), some of the LCS clearly disappeared while others preserved. However, as in the temporal scale 1, the web-like LCS were preserved in the spatial scales 2 and 3, and lost in some regions of the spatial scale 4 (bottom panel; second left column). In the temporal scale 3 (second right column), further disappearance of LCS was observed in the spatial scales 1 and 2. However, the web-like LCS were still preserved in the spatial scales 3. The attracting LCS were lost in the spatial scale 4 (bottom panel; second right column). In the temporal scale 4 (right column), LCS disappeared further but some linear structures remained in the spatial scales 1, 2 and 3. As in the temporal scale 3,  the attracting LCS were lost in the spatial scale 4 (bottom panel; right column).

\subsection{Scale-invariant Lagrangian coherent structures.}
\begin{figure}
  \centering
  \includegraphics[width=0.8\linewidth,trim={0cm 0cm 0cm 0cm},clip]{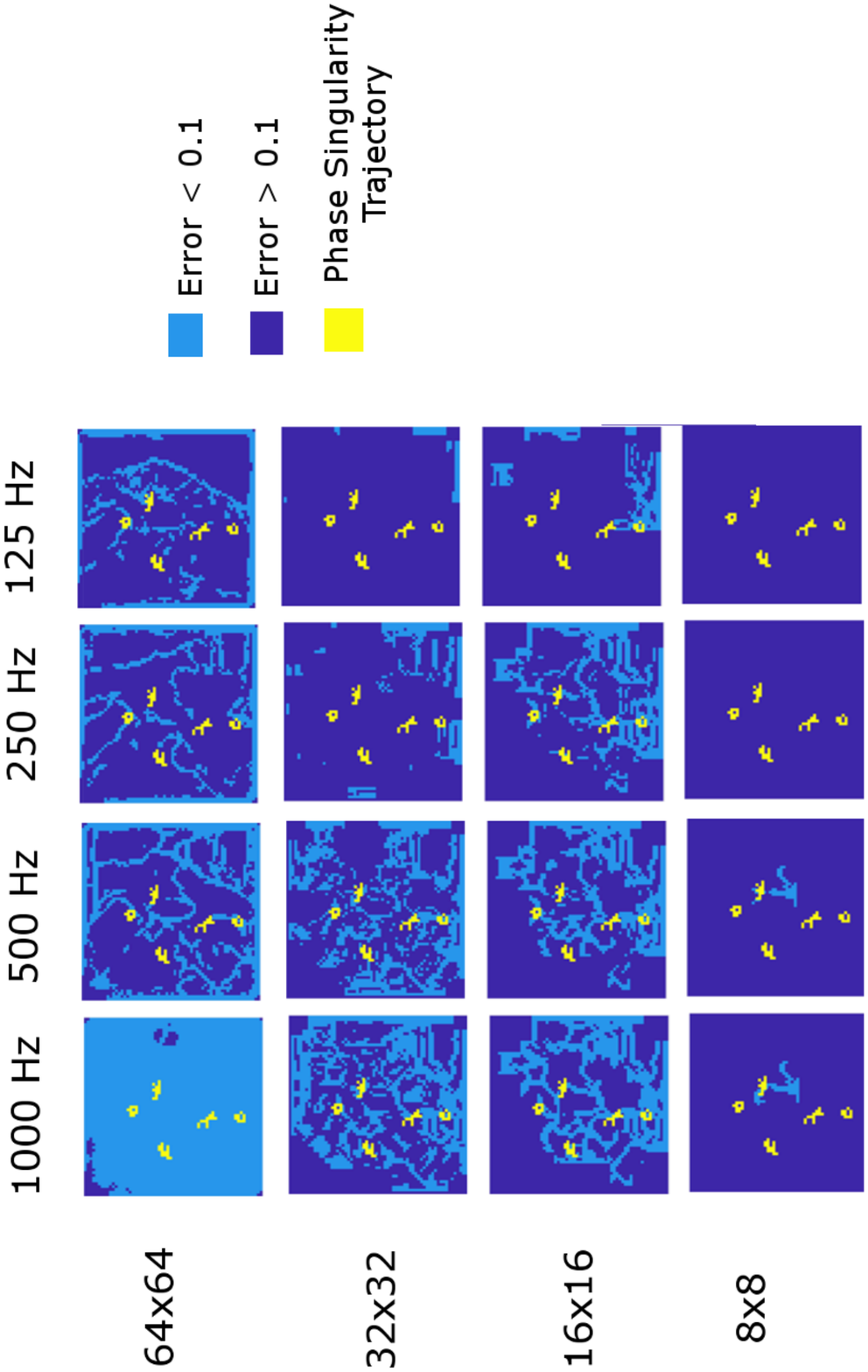}
  \caption{
    \textbf{Assessment of scale-invariance of repelling Lagrangian coherent structures.} The renormalization group of a total of 16 (= $4 \times 4$) spatiotempral scales. The $x$-axis is the temporal scale (1 through 4) and the $y$-axis is the spatial scale (1 through 4). The components shown in light blue indicate those with small finite-time Lyapunov exponent (FTLE) errors ($<$ 0.1). The components shown in dark blue indicate those with large finite-time Lyapunov exponent (FTLE) errors ($\geq$ 0.1). The yellow lines represent the trajectory of the rotors.
  }
  \label{fig:repelling}
\end{figure}

\begin{figure}
  \centering
  \includegraphics[width=0.8\linewidth,trim={0cm 0cm 0cm 0cm},clip]{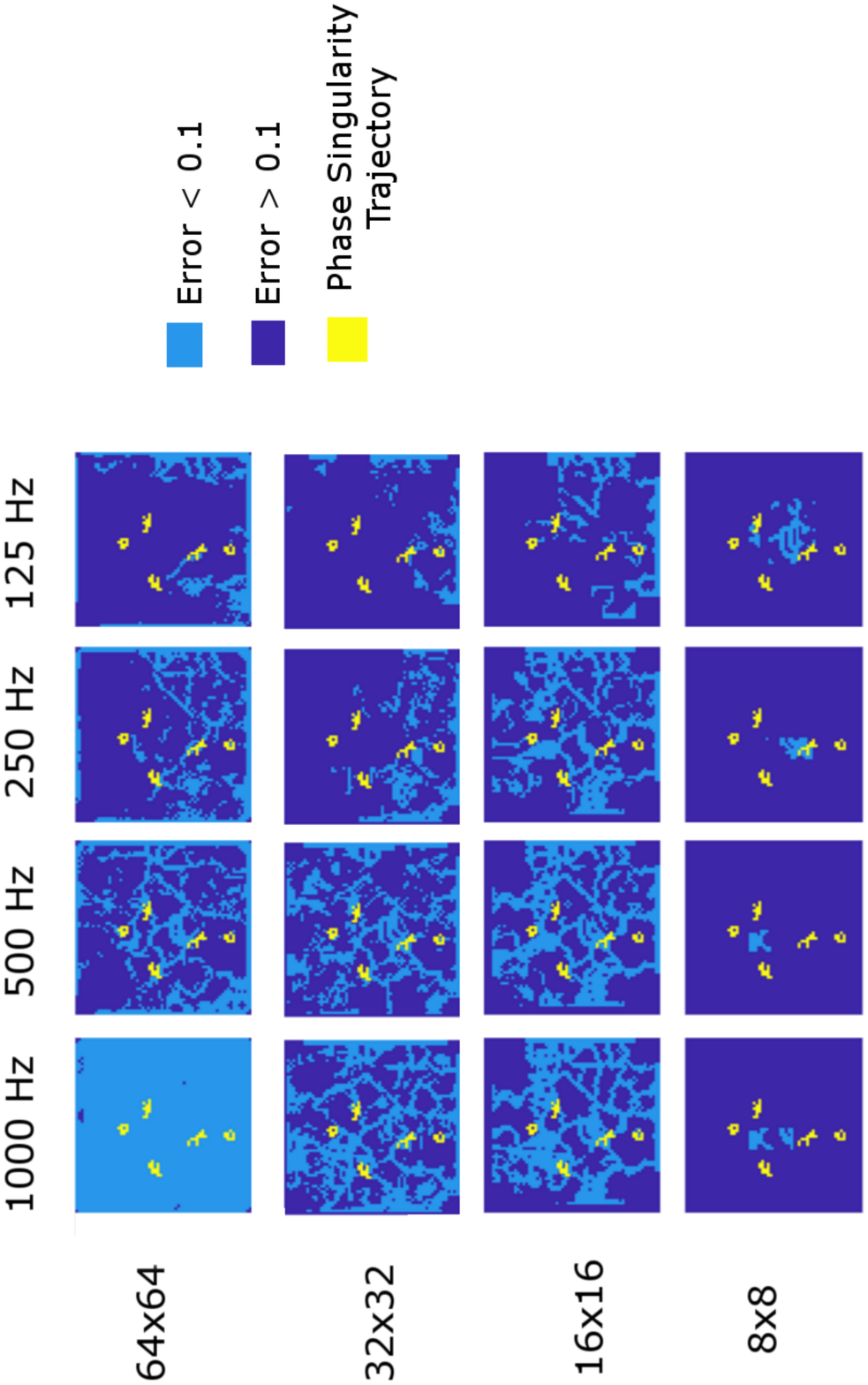}
  \caption{
    \textbf{Assessment of scale-invariance of attracting Lagrangian coherent structures.} The renormalization group of a total of 16 (= $4 \times 4$) spatiotempral scales. The $x$-axis is the temporal scale (1 through 4) and the $y$-axis is the spatial scale (1 through 4). The components shown in light blue indicate those with small finite-time Lyapunov exponent (FTLE) errors ($<$ 0.1). The components shown in dark blue indicate those with large finite-time Lyapunov exponent (FTLE) errors ($\geq$ 0.1). The yellow lines represent the trajectory of the rotors.
  }
  \label{fig:attracting}
\end{figure}

In the temporal scale 1 (left column), preservation of the repelling LCS was observed in the spatial scale 2 and 3 (Figure \ref{fig:repelling}). However, the repelling LCS was virtually lost in the spatial scale 4 (bottom left panel). Similar observations were made in the temporal scale 2 (second left column) and 3 (second right column). In the temporal scale 4 (right column), preservation of the repelling LCS was only observed in the spatial scale 1, whereas it was virtually lost in the spatial scale 3 and 4. 
For the attracting LCS, preservation of LCS was observed in the spatial scale 2 and 3 of the temporal scale 1 (left column; Figure \ref{fig:attracting}). Similar to the repelling LCS, however, the attracting LCS was virtually lost in the spatial scale 4 (bottom left panel). Similar observations were made in the temporal scale 2 (second left column) and 3 (second right column). In the temporal scale 4 (right column), preservation of the attracting LCS was only observed in the spatial scale 1, whereas it was virtually lost in the spatial scale 3 and 4. Importantly, those scale-invariant LCS were not co-localized to the trajectory of the rotors (Figure 3 and 4). 

\section{Discussion}
\subsection{Main Findings}
Our main findings are summarized as follows. First, we found that both the repelling and attracting LCS change across the different spatial and temporal scales of the renormalization group. Second, despite the change across the scales, some LCS are scale-invariant, particularly down to the spatial and temporal scales 3. Third, the patterns of those scale-invariant structures are not obvious from the trajectory of the rotors based on the traditional voltage mapping of the cardiac system. 

\subsection{Lagrangian coherent structures of the cardiac system}
The repelling LCS of information flow indicates a surface barrier that separates the individual information
flow. In other words, the repelling LCS segment the lattice into smaller segments of information dynamics. In
contrast, the attracting LCS of information flow represents a region of information mixing, which can be considered as a meeting point of information particles that originate from different spiral waves. Our approach provides a tool to quantitatively characterize a macro-scale behavior of excitable media by specifically focusing on information transport, thereby quantifying the spiral wave dynamics. In our previous work, we applied our method to a simple model of excitable media to quantify the complex interactions among the components \cite{ashikaga2017hidden}. The present work demonstrates the applicability of our method to a cardiac system to quantify the information flow underlying spiral wave dynamics. 

\subsection{Clinical implications}
Our findings have two important clinical implications. First, our analysis sheds new light on the mechanism that maintains fibrillation. Our findings indicate the presence of scale-invariant structures associated with spiral wave dynamics. It is possible that those scale-invariant structures quantified by the LCS contributes to maintenance of spiral waves. Second, our analysis provides a new approach to quantifying fibrillation, rather than simply the presence or absence of fibrillation. Our method of quantitative analysis of human fibrillation provides patient-specific diagnostic parameters that could potentially serve as a valid endpoint for therapeutic interventions.

\subsection{Limitations}
We recognize several limitations associated with the numerical method we implemented. We used a Fenton-Karma model, which is a relatively simple cardiac model, with a homogeneous and isotropic 2-D lattice. It is possible that a more biophysically detailed model of the heart with anatomical heterogeneity, anisotropy and a more realistic geometry could make our approach more difficult to analyze. However, the information-theoretic approach that we used in the study are independent of any specific trajectory of each spiral wave. In addition, the simplicity of the cardiac model is an advantage that allows the results from this model to be widely applicable to other systems. 

\section{Conclusions}
The Lagrangian coherent structures of information flow underlying spiral waves are preserved across multiple spatiotemporal scales. A multi-scale approach to information flow within the cardiac system provides a quantitative tool to improve our understanding of the mechanism of fibrillation.



\section{References}
\bibliographystyle{elsarticle-num} 
\bibliography{fib_ref}

\begin{thebibliography}{10}
\expandafter\ifx\csname url\endcsname\relax
  \def\url#1{\texttt{#1}}\fi
\expandafter\ifx\csname urlprefix\endcsname\relax\def\urlprefix{URL }\fi
\expandafter\ifx\csname href\endcsname\relax
  \def\href#1#2{#2} \def\path#1{#1}\fi

\bibitem{Lip2016aa}
G.~Y.~H. Lip, L.~Fauchier, S.~B. Freedman, I.~Van~Gelder, A.~Natale, C.~Gianni,
  S.~Nattel, T.~Potpara, M.~Rienstra, H.-F. Tse, D.~A. Lane,
  \href{http://dx.doi.org/10.1038/nrdp.2016.16}{Atrial fibrillation}, Nat Rev
  Dis Primers 2 (2016) 16016.
\newline\urlprefix\url{http://dx.doi.org/10.1038/nrdp.2016.16}

\bibitem{haissaguerre2014driver}
M.~Haissaguerre, M.~Hocini, A.~Denis, A.~J. Shah, Y.~Komatsu, S.~Yamashita,
  M.~Daly, S.~Amraoui, S.~Zellerhoff, M.-Q. Picat, et~al., Driver domains in
  persistent atrial fibrillation, Circulation 130~(7) (2014) 530--8.

\bibitem{benharash2015quantitative}
P.~Benharash, E.~Buch, P.~Frank, M.~Share, R.~Tung, K.~Shivkumar, R.~Mandapati,
  Quantitative analysis of localized sources identified by focal impulse and
  roter modulation mapping in atrial fibrillation, Circ Arrhythm Electrophysiol
  8~(3) (2015) 554--61.

\bibitem{gianni2016acute}
C.~Gianni, S.~Mohanty, L.~Di~Biase, T.~Metz, C.~Trivedi, Y.~G{\"o}ko{\u{g}}lan,
  M.~F. G{\"u}ne{\c{s}}, R.~Bai, A.~Al-Ahmad, J.~D. Burkhardt, et~al., Acute
  and early outcomes of focal impulse and rotor modulation (firm)-guided
  rotors-only ablation in patients with nonparoxysmal atrial fibrillation,
  Heart Rhythm 13~(4) (2016) 830--835.

\bibitem{berntsen2016focal}
R.~F. Berntsen, T.~F. H{\aa}land, R.~Sk{\aa}rdal, T.~Holm, Focal impulse and
  rotor modulation as a stand-alone procedure for the treatment of paroxysmal
  atrial fibrillation: A within-patient controlled study with implanted cardiac
  monitoring, Heart Rhythm 13~(9) (2016) 1768--1774.

\bibitem{buch2016long}
E.~Buch, M.~Share, R.~Tung, P.~Benharash, P.~Sharma, J.~Koneru, R.~Mandapati,
  K.~A. Ellenbogen, K.~Shivkumar, Long-term clinical outcomes of focal impulse
  and rotor modulation for treatment of atrial fibrillation: A multicenter
  experience, Heart Rhythm 13~(3) (2016) 636--641.

\bibitem{ashikaga2017hidden}
H.~Ashikaga, R.~G. James, Hidden structures of information transport underlying
  spiral wave dynamics, Chaos 27~(1) (2017) 013106.

\bibitem{haller2000lagrangian}
G.~Haller, G.~Yuan, Lagrangian coherent structures and mixing in
  two-dimensional turbulence, Physica D: Nonlinear Phenomena 147~(3-4) (2000)
  352--370.

\bibitem{shadden2005definition}
S.~C. Shadden, F.~Lekien, J.~E. Marsden, Definition and properties of
  lagrangian coherent structures from finite-time lyapunov exponents in
  two-dimensional aperiodic flows, Physica D 212~(3) (2005) 271--304.

\bibitem{haller2015lagrangian}
G.~Haller, Lagrangian coherent structures, Annu Rev Fluid Mech 47 (2015)
  137--162.

\bibitem{kadanoff1966scaling}
L.~P. Kadanoff, Scaling laws for ising models near tc, Physics 2~(6) (1966)
  263--272.

\bibitem{ashikaga2018causal}
H.~Ashikaga, F.~Prieto~Castrillo, M.~Kawakatsu, N.~Dehghani, Causal scale of
  rotors in a cardiac system, Front Phys 6 (2018) 30.

\bibitem{ashikaga2018inter}
H.~Ashikaga, R.~G. James, Inter-scale information flow as a surrogate for
  downward causation that maintains spiral waves, Chaos: An Interdisciplinary
  Journal of Nonlinear Science In Press.

\bibitem{fenton1998vortex}
F.~Fenton, A.~Karma, Vortex dynamics in three-dimensional continuous myocardium
  with fiber rotation: Filament instability and fibrillation, Chaos: An
  Interdisciplinary Journal of Nonlinear Science 8~(1) (1998) 20--47.

\bibitem{fenton2002multiple}
F.~H. Fenton, E.~M. Cherry, H.~M. Hastings, S.~J. Evans, Multiple mechanisms of
  spiral wave breakup in a model of cardiac electrical activity, Chaos 12~(3)
  (2002) 852--92.

\bibitem{pertsov1993spiral}
A.~M. Pertsov, J.~M. Davidenko, R.~Salomonsz, W.~T. Baxter, J.~Jalife, Spiral
  waves of excitation underlie reentrant activity in isolated cardiac muscle.,
  Circ Res 72~(3) (1993) 631--50.

\bibitem{ashikaga2018locating}
H.~Ashikaga, A.~Asgari-Targhi, Locating order-disorder phase transition in a
  cardiac system, Sci Rep 8~(1) (2018) 1967.

\bibitem{aronis2018impact}
K.~N. Aronis, H.~Ashikaga, Impact of number of co-existing rotors and
  inter-electrode distance on accuracy of rotor localization, J Electrocardiol
  51 (2018) 82--91.

\bibitem{schreiber2000measuring}
T.~Schreiber, Measuring information transfer, Phys Rev Lett 85~(2) (2000) 461.

\bibitem{lizier2014jidt}
J.~T. Lizier, Jidt: An information-theoretic toolkit for studying the dynamics
  of complex systems, Front Robot AI 1 (2014) 11.

\end{thebibliography}





\end{document}